\documentclass[pra,twocolumn,amsmath, amssymb]{revtex4}
\usepackage{graphicx}
\usepackage{dcolumn}
\usepackage{amsmath}
\usepackage{epsfig}
\usepackage{amsfonts,amssymb}
\begin{document}
\title{Hyperentanglement-enhanced Weak Value Amplification with High-Order Correlation Function}
\author{Jing-Zheng Huang\footnote[1]{These authors contributed equally to this work.}\footnote[2]{jzhuang1983@sjtu.edu.cn}, Qin-Zheng Li*, Chen Fang and Gui-Hua Zeng\footnote[3]{ghzeng@sjtu.edu.cn}}%
\affiliation{State Key Laboratory of Advanced Optical Communication Systems and Networks, and
Center of Quantum Information Sensing and Processing, Shanghai Jiaotong University, Shanghai 200240, China}

\begin{abstract}   
By exploiting the hyperentanglement, we propose a new concrete scheme to realize weak value amplification, which can be implemented by $N$ hyperentangled photons along with coincidence detection. Our scheme can optimally increase the amplification factor by a factor of $N$ compared to the usage of $N$ uncorrelated photons, and achieve the Heisenberg limit for the purpose of parameter estimation.
We expect our study can inspire a further investigation on the usage of hyperentanglement in weak measurement and quantum metrology.
\end{abstract}
 \maketitle


\section*{\uppercase\expandafter{\romannumeral1}. Introduction}
As an advanced technique that provides higher sensitivity for parameter estimation, the weak value amplification (WVA) method has attracted much attention in the recent years\cite{Aharonov1988,Dixon2009,Hosten2008,Starling2010,Xu2013,Starling20102,Loaiza2014,Malik2014,Brunner2010,Strubi2013,Brunner2004,Viza2013}.
Of which the fundamental advantage is still in debate\cite{Knee2014,Ferrie2014}, WVA is shown useful in overcoming certain types of technical noises\cite{Jordan2014}, and in some situations outperforms the standard interferometric techniques\cite{Brunner2010,Dressel2014}.

However, the low successful probability of the postselection process appears to be a shortcoming of WVA. Recently, Pang, Dressel and Brun\cite{Pang2014} show that the scheme efficiency can be highly improved by utilizing quantum entanglement. In their proposal, entangled ancillary systems are prepared and postselected, by doing so one can increase the postselection probability with the amplification factor preserved, or alternatively enhance the amplification factor with the post-selection possibility preserved. As both of these enhancements scale optimally with the entangled ancillas number, either of the two cases can achieve the Heisenberg scaling of estimation precision, as is expected in quantum metrology\cite{Giovannetti2004}.

In this paper, we propose a new scheme to realized WVA, by exploiting a phenomenon called "hyperentanglement"\cite{Kwiat1997}, which exists in various physical systems.
By first using first-order correlation to explain the WVA phenomenon, we generalize this theory to high-order scenario and then devise the usage of hyperentanglement. We show that while a maximum success probability similar to \cite{Pang2014} is achieved, our scheme can also optimally increase the amplification factor by a factor of $N$, compared to the optimal scale of $\sqrt{N}$ in \cite{Pang2014}. Interestingly, it is the real part of weak value maximizes the amplification factor, in contrast to the use of the imaginary part in many previous studies including \cite{Pang2014}. Although in this case the success probability is irrelevant to $N$, the the higher amplification factor makes the Heisenberg limit for parameter estimation still achievable in our scheme. More importantly, our proposal indicates the advances of involving hyperentanglement to weak measurement, and for the first time reveals the relation between WVA phenomenon and the correlation functions of field, which opens a new way for a further investigation in this area.

This paper is organized as follows: In Sec.\uppercase\expandafter{\romannumeral2}, we briefly review and describe the standard weak value amplification process in the theory of first-order correlation function. In Sec.\uppercase\expandafter{\romannumeral3}, we propose our hyperentanglement-enhanced proposal and generalize the above theory to high-order scenarios. A followed Fisher information analysis derives our claims. Discussions and conclusions are summarized in Sec.\uppercase\expandafter{\romannumeral4}.

\section*{\uppercase\expandafter{\romannumeral2}. Preliminaries}

Weak measurement can be described by a Hamiltonian with the form of
\begin{equation}\label{eq:Hermitian}
H=g\delta(t-t_0)\hat{A}\otimes\hat{M}.
\end{equation}
Here, the Dirac delta function $\delta(t-t_0)$ represents an impulsive interaction at $t=t_0$, corresponding to an unitary evolution as $\hat{U}_g=exp[-ig\hat{A}\otimes\hat{M}]$ (we have set $\hbar=1$ without loss of generality); $\hat{A}$ and $\hat{M}$ are two operators corresponding to the ancillary system and the meter respectively, it is convenient to set $\hat{A}$ as the Z operator and $\hat{M}$ as the position or momentum operator for optical experiments\cite{Brunner2010,Xu2013}; $g$ represents the coupling parameter, which is extremely small in the context of weak measurement.

Denoting the initial and postselected states of the system as $|i\rangle$ and $|f\rangle$ respectively, the \emph{weak value} about $\hat{A}$ is defined as $A_w \equiv \langle f|\hat{A}|i\rangle/\langle f|i\rangle$\cite{Aharonov1988}. After a successful postselection on the system, one can observe an anomalous detector response on the meter. For example by setting $\hat{M}=\hat{P}$ the average value of $\hat{X}$(position) or $\hat{P}$(momentum) yield $\langle\hat{X}\rangle_f = \langle\hat{X}\rangle_i + g\mathrm{Re}A_w$ or $\langle\hat{P}\rangle_f = \langle\hat{P}\rangle_i - 2g\mathrm{Im}A_w Var(\hat{P})_i$, where $Re/Im$ denotes the real/imaginary part of a complex number and $Var()$ denotes the variance of the operator in the bracket\cite{Jozsa2007}.
The value of $g$ can be estimated from the displacement, with an amplification factor of $\mathrm{Re}A_w$(measuring $\langle\hat{X}\rangle$) or $2\mathrm{Im}A_w Var(\hat{P})_i$(measuring $\langle\hat{P}\rangle$).

These results can be alternatively derived from the theory of first-order correlation function\cite{Scully1997}. First, we write the initial total state as $|\Phi_i\rangle = |\Psi_i\rangle\otimes|\varphi_i\rangle = (\int dx \psi(x)a^{\dag}_{x}|0\rangle)\otimes|\varphi_i\rangle$ (or $(\int dp  \widetilde{\psi}(p)a^{\dag}_{p}|0\rangle)\otimes|\varphi_i\rangle$, depends on the expansion basis one chooses).
Here, $|\Psi_i\rangle$ represents the initial meter state, $a^{\dag}_x$ (or $a^{\dag}_p$) is the creation operator on the vacuum state $|0\rangle$, generates a photon or particle in position mode (or momentum mode) with probability of $|\psi(x)|^2$(or $|\widetilde{\psi}(p)|^2$);
$|\varphi_i\rangle$ represents the initial system state, which can be expanded by $=\sum_{i}c_i|a_i\rangle$ on the basis of $\{|a_i\rangle\}$\cite{Note}.

Followed the unitary evolution described by $\hat{U}_g$, one can postselect the system by a final state $|\varphi_f\rangle$ (see fig.~\ref{fig:second}(a)). After this postselection, the meter state collapses to be\cite{Jozsa2007}:
\begin{equation}\label{eq:final}
\begin{aligned}
|\Psi_f\rangle =& \langle\varphi_f|\hat{U}|\Phi_i\rangle
\approx \langle\varphi_f|\mathbb{I}-ig\hat{A}\hat{M}|\Phi_i\rangle\\
=&\int dx \psi(x)a^{\dag}_{x}|0\rangle\langle\varphi_f|\varphi_i\rangle\\
&-i\hat{M}\int dx g\psi(x)a^{\dag}_{x}|0\rangle\langle\varphi_f|\hat{A}|\varphi_i\rangle\\
&= \langle\varphi_f|\varphi_i\rangle(1-igA_w\hat{M})\int dx \psi(x)a^{\dag}_{x}|0\rangle\\
&\approx \langle\varphi_f|\varphi_i\rangle e^{-igA_w\hat{M}}\int dx \psi(x)a^{\dag}_{x}|0\rangle\\
&= \langle\varphi_f|\varphi_i\rangle e^{-igA_w\hat{M}}|\Psi_i\rangle.
\end{aligned}
\end{equation}
where we expand it on the $\{|x\rangle\}$ basis. The successful probability of the postselection is $p_s = |\langle\varphi_f|\varphi_i\rangle|^2$.

To illustrate the derivation more clearly, let us from now on restrict our discussions on optics. In other word, the system state $|\varphi\rangle$ is chosen to be a state of photon polarization, and the meter is chosen to be the arrival time (corresponding to position $\hat{X}$) or optical frequency (corresponding to momentum $\hat{P}$) of photon\cite{Brunner2010}. In this context, the detector of the meter can be described by the electrical field operator:
\begin{equation}\label{eq:field}
\hat{E}^{\dag}(p)=\int dx E_0e^{-ipx}\hat{a}_{x}.
\end{equation}
Note that for photons $x$ and $p$ can be substituted by $ct$ and $\hbar\omega/c$, thus the detector records the optical frequencies of the arrival photons. Let $\hat{M}=\hat{X}$ and apply the commutation relation $[a_{x'}, a^{\dag}_{x}]=\delta(x'-x)$, the first-order self correlation function can be derived as:
\begin{eqnarray}\label{eq:reorder}
\begin{aligned}
G^{(1)}(p) &= \langle \Psi_f|E^-(p)E^+(p)|\Psi_f\rangle = |\langle 0|E^+(p)|\Psi_f\rangle|^2 \\
           &= p_s|E_0|^2|\int dx e^{-ixp} e^{-igA_wx} \psi(x)|^2\\
           &\propto |\int dx e^{-i(p+g\mathrm{Re}A_w)x}e^{g\mathrm{Im}A_w x}e^{-x^2\sigma^2}|^2\\
           &\propto \exp[-\frac{(p+g\mathrm{Re}A_w)^2}{2\sigma^2}],
\end{aligned}
\end{eqnarray}
where we assume that $|\psi(x)|^2$ obeys Gaussian distribution with zero mean value and variance of $\sigma^2_x = 1/4\sigma^2$ (which implies the variance of $|\widetilde{\psi}(x)|^2$ is $\sigma^2_p = \sigma^2$). There is a displacement on the average value of $p$ of $g\mathrm{Re}A_w\sigma^2_p$, which is coincident to the standard WVA theory\cite{Jozsa2007}.

The detector can also record the arrival time of the photons, in this case we should calculate the first-order self-correlation function about $x$. Similar derivations give that:
\begin{eqnarray}\label{eq:reorder}
\begin{aligned}
G^{(1)}(x) \propto \exp[-\frac{(x+2g\mathrm{Im}A_w\sigma_x^2)^2}{2\sigma_x^2}],
\end{aligned}
\end{eqnarray}
which also agrees with the standard theory.

\section*{\uppercase\expandafter{\romannumeral3}. Hyperentanglement-enhanced Weak Value Amplification}

\subsection*{\uppercase\expandafter{\romannumeral3.A}. The Scheme}

\begin{figure}[!h]\center
\resizebox{8cm}{!}{
\includegraphics{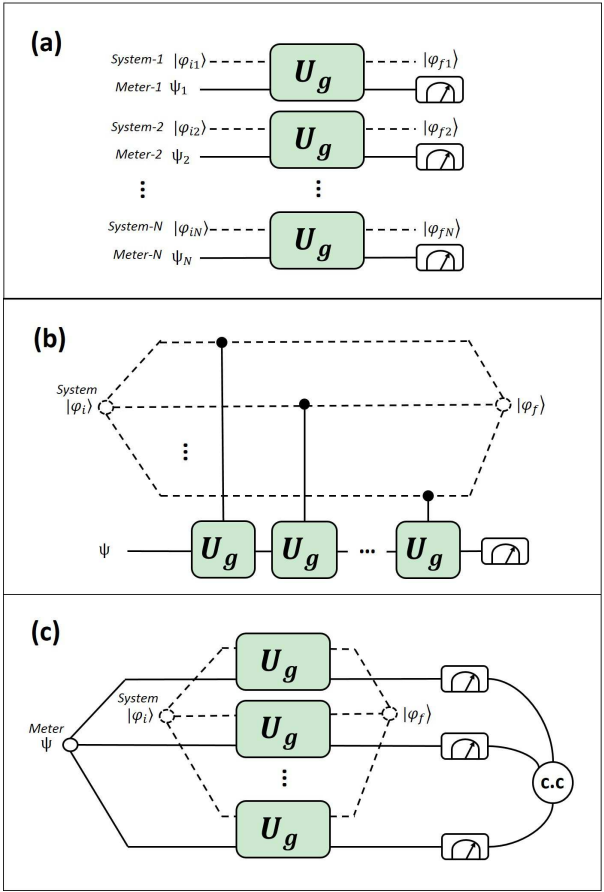}}
\caption{Comparison of three types of weak value amplification(WVA) schemes.
(a) In the standard WVA scheme, $N$ uncorrelated attempts between ancillary systems and meters are performed.
(b) In the entanglement-assisted WVA\cite{Pang2014}, $N$ entangled ancillary systems sequentially interact with a meter, parameter information is read out from the detector after a successful entangled postselection on the ancillas.
(c) In the hyperentanglement-assisted WVA, parallel interactions between $N$ entangled ancillary systems and $N$ entangled meter are performed, parameter information is read out from the coincidence detection on the meters after a successful entangled postselection on the ancillas.}\label{fig:second}
\end{figure}

The relationship between WVA and first-order correlation function can be generalized to the higher order scenarios when hyperentangled photons are involved. Before developing this theory, we first introduce our hyperentanglement-assisted WVA scheme depicted in fig.~\ref{fig:second}(c). In this scheme, $N$ entangled photons are generated, by e.g. parametric down conversion process. These photons are simultaneously entangled in polarizations and momentums(or, alternatively, optic frequencies), they respectively experience unitary evolution described by $\hat{U}_g = \exp(-ig\hat{A}\hat{M})$, where $\hat{A}=|H\rangle\langle H|-|V\rangle\langle V|$ and $\hat{M}$ the position operator $\hat{X}$ or momentum operator $\hat{P}$. Making the polarizations as the systems and the photon momentums as the meters, the photons after evolution are postselected in polarization by $|\psi_f\rangle$. Coincidence detection are performed by recording the optic frequencies of every arrival photons, which reveals the $N^{th}$-order correlation function\cite{Scully1997}. Similar to the first-order case, WVA phenomenon can be observed from $N^{th}$-order correlation function, where the theory will be fully discussed in Sec.\uppercase\expandafter{\romannumeral3.B}.

For a comparison, the entanglement-assisted scheme proposed by ref.\cite{Pang2014} is depicted in fig.~\ref{fig:second}(b).
Our scheme is different from it in three aspects. First, instead of using a single meter, we apply entangled meters with the same number of the ancillas. Thanks to the advanced feature of hyperentanglement\cite{Barreiro2005}, correlated photons produced by a nonlinear optical process are simultaneously entangled in multiple degrees of freedom, such as their polarizations and momentums\cite{Barbieri2005,Gao2010}. By exploiting the polarizations as ancillary systems and the momentums as meters, our scheme can be implemented by optical components. Second, instead of sequentially interacting the meter with $N$ ancillas, parallel one-on-one interactions are performed between the $N$ pairs of meters and ancillas in our scheme. Finally, instead of detecting the state of one meter, our scheme reveals the parameter information from the high order correlation function of $N$ meters by performing coincidence detection on them.

The advantage of our scheme is twofold:

(1) Compared to the standard WVA proposal\cite{Dressel2014}, our scheme can simultaneously get increased amplification factor and higher success probability. As a consequence, Heisenberg scaling precision in parameter estimation can be achieved. Detailed calculations supporting these claims will be presented in Sec.\uppercase\expandafter{\romannumeral3.B} and Sec.\uppercase\expandafter{\romannumeral3.C}.

(2) Compared to the entanglement-assisted WVA scheme, our scheme can at most increase the weak value by a factor of $N$, rather than $\sqrt{N}$ in \cite{Pang2014}. Moreover, our scheme also advantages in the experimental realization. Here we consider using photons as carriers because they are much easier to manipulate than atomic particles. To realize the proposal in Ref.\cite{Pang2014}, it requires weak coupling interactions between photons which are hard to control\cite{Shapiro2006}. Our scheme avoids this obstacle, because the weak coupling interactions between two freedom degrees happen independently on each photon, which are easy to manipulate as shown in plenty of previous experiments\cite{Hosten2008,Dixon2009,Starling2010,Xu2013}.

\subsection*{\uppercase\expandafter{\romannumeral3.B}. The Theory}

To derive the most general theory, we should start by studying the special case of $N = 2$. The initial state of a hyperentangled photon pair, which is usually generated by the spontaneous parametric down conversion (SPDC) process\cite{Howell2004}, is described as:
\begin{subequations}\label{eq:pres}
\begin{align}
|\Phi_i\rangle &\sim [\int\int dx_1dx_2 \psi(x_1,x_2)a^{\dag}_{x_1}a^{\dag}_{x_2}|0,0\rangle]\otimes|\varphi_i\rangle \nonumber\\
or,\nonumber\\
               &\sim [\int\int dp_1dp_2 \widetilde{\psi}(p_1,p_2)a^{\dag}_{p_1}a^{\dag}_{p_2}|0,0\rangle]\otimes|\varphi_i\rangle. \nonumber
\end{align}
\end{subequations}
For type-I nondegenerate and type-II SPDC, the probability amplitude has the form of $\widetilde{\psi}(p_1,p_2) \propto \exp[-\frac{(p_1+p_2-p_0)^2}{4\sigma_0^2}]\cdot\frac{\sin[D(p_1-p_2)]}{D(p_1-p_2)}$, with $p_0$ and $\sigma_0$ corresponding to the mean value and variance of the pump light momentum and D a constant coefficient\cite{Valencia2002}, reflecting the entanglement on the photon momentum (and thus the arrival time). On the other hand, $|\varphi_i\rangle$ the system state is also entangled in polarization: $|\varphi_i\rangle=\frac{|HH\rangle+|VV\rangle}{\sqrt{2}}$, with $H$ and $V$ stand for two orthogonal polarization directions.

The interaction on systems and meters for these two photons is $\hat{U}=\hat{U}_{g1}\otimes\hat{U}_{g2} = \exp[-ig\hat{A}_1\hat{M}_1]\otimes\exp[-ig\hat{A}_2\hat{M}_2]$.
By postselecting the system with a final state $|\varphi_f\rangle=\frac{1}{\sqrt{2}}(e^{-i\epsilon}|HH\rangle-e^{i\epsilon}|VV\rangle)$, the final meter state yields:
\begin{equation*}\label{eq:entfinal}
\begin{aligned}
|\Psi_{fM}\rangle =&\langle\varphi_f|\hat{U}|\Psi_i\rangle\\
               =&\langle\varphi_f|(\mathbb{I}_1\otimes\mathbb{I}_2-ig\hat{A}_1\hat{M}_1\otimes\mathbb{I}_2-\mathbb{I}_1\otimes ig\hat{A}_2\hat{M}_2+\cdots)|\Psi_i\rangle\\
               \approx& \langle\varphi_f|\varphi_i\rangle\{\int\int dx_1 dx_2\psi(x_1,x_2)\times\\
               &(1-ig\frac{\langle\varphi_f|\hat{A}_1\otimes \mathbb{I}_2|\varphi_i\rangle}{\langle\varphi_f|\varphi_i\rangle}\hat{M}_1-
               ig\frac{\langle\varphi_f|\mathbb{I}_1 \otimes\hat{A}_2|\varphi_i\rangle}{\langle\varphi_f|\varphi_i\rangle}\hat{M}_2)\times\\
               &a^{\dag}_{x_1}a^{\dag}_{x_2}|0,0\rangle\}.\\
\end{aligned}
\end{equation*}
In this case, the postselection probability is $p_s=|\langle\varphi_f|\varphi_i\rangle|^2=|-i\sin(\epsilon)|^2\approx\epsilon^2$, and the the weak value divides into two parts: $A_{w1} = \frac{\langle\varphi_f|\hat{A}_1\otimes \mathbb{I}_2|\varphi_i\rangle}{\langle\varphi_f|\varphi_i\rangle}$ and $A_{w2}=\frac{\langle\varphi_f|\mathbb{I}_1\otimes \hat{A}_2|\varphi_i\rangle}{\langle\varphi_f|\varphi_i\rangle}$. Applying the calculating skill in Sec.\uppercase\expandafter{\romannumeral2}, we can compute the second-order correlation function about $p$ for $\hat{M}=\hat{X}$ and $\hat{M}=\hat{P}$:
\begin{subequations}\label{eq:entim}
\begin{align}
G^{(2)}_{\hat{X}}(p_1,p_2) =& |\langle0,0|E^+(p_1)E^+(p_2)|\Psi_{fX}\rangle|^2 \nonumber\\
\propto& \exp{\{-\frac{[p_1+p_2-p_0+g\mathrm{Re}(A_{w1}+A_{w1})]^2}{2\sigma_0^2}\}}\\
G^{(2)}_{\hat{P}}(p_1,p_2) =& |\langle0,0|E^+(p_1)E^+(p_2)|\Psi_{fP}\rangle|^2 \nonumber\\
\propto& \exp{\{-\frac{[p_1+p_2-p_0+2g(\mathrm{Im}\overline{A_{w}})\sigma^2]^2}{2\sigma^2}\}}.
\end{align}
\end{subequations}
In eq.(\ref{eq:entim}b), we set $\mathrm{Im}A_{w1}=\mathrm{Im}A_{w2}\equiv\mathrm{Im}\overline{A_{w}}$ for simplicity.
From above we can notice that for $\hat{M}=\hat{X}$, the weak value is doubled compared to the first-order case, while this effect does not exist for $\hat{M}=\hat{P}$. Interestingly, real weak value introduces a bigger amplification factor, in contrast to the first-order case where imaginary weak value is preferred.

Eq.(\ref{eq:entim}) can be easily generalized to the case of $N$ hyperentangled photons are in use, where we can obtain the $N^{th}$-order correlation function as:
\begin{subequations}\label{eq:GNent}
\begin{align}
G^{(N)}_{\hat{X}}(\{p_n\}) &\propto \exp{(-(\sum_{n=1}^{N}p_n-p_0+g\mathrm{Re}A_w)^2/2\sigma_0^2)},\\
G^{(N)}_{\hat{P}}(\{p_n\}) &\propto \exp{(-(\sum_{n=1}^{N}p_n-p_0+2g\mathrm{Im}\overline{A_w}\sigma_0^2)^2/2\sigma_0^2)},
\end{align}
\end{subequations}
where we similarly set $A_{wn} = \overline{A_w}$ for $(n=1,...,N)$ and denote $A_w \equiv \sum_{n=1}^{N}A_{wn}$.

For some simple examples, given
\begin{equation}
\begin{aligned}\label{eq:i-f}
|\psi_i\rangle &= \frac{1}{\sqrt{2}}(|H\rangle^{\otimes N}+ |V\rangle^{\otimes N})~~and\\
|\psi_f\rangle &= \cos(-\frac{\pi}{4}+\epsilon)|H\rangle^{\otimes N}+ \sin(-\frac{\pi}{4}+\epsilon)|V\rangle^{\otimes N},
\end{aligned}
\end{equation}
the setting of $\hat{M}=\hat{X}$ reveals a corresponding weak value of $A_w \approx \frac{N}{\epsilon}$, which enhances by a factor of $N$ compared to the optimal scale of $\sqrt{N}$ in \cite{Pang2014}.
Moreover, if $|\psi_i\rangle = \frac{1}{\sqrt{2}}(|H\rangle^{\otimes N}+ |V\rangle^{\otimes N})$ and $|\psi_f\rangle = \cos(-\frac{\pi}{4}+N\epsilon)|H\rangle^{\otimes N}+ \sin(-\frac{\pi}{4}+N\epsilon)|V\rangle^{\otimes N}$, with the weak value maintains the same as using $N$ uncorrelated photons, the probability of a success postselection becomes $p_s \approx N^2\epsilon^2$, which matches the optimal probability derived in \cite{Pang2014}.
More generally, by setting the final state to be $|\psi_f\rangle = \cos(-\frac{\pi}{4}+k\epsilon)|H\rangle^{\otimes N}+ \sin(-\frac{\pi}{4}+k\epsilon)|V\rangle^{\otimes N}$ with $1 < k < N$, we have $p_s \approx k^2\epsilon^2$ and $A_w \approx \frac{N}{k\epsilon}$, which means one can simultaneously get higher success probability and larger amplification factor than standard WVA without assistance of entanglement or hyperentanglement.

Meanwhile, if $|\psi_f\rangle = \frac{1}{\sqrt{2}}(e^{-i\epsilon}|H\rangle^{\otimes N}+ e^{-i\epsilon}|V\rangle^{\otimes N})$, the setting of $\hat{M}=\hat{P}$ reveals an average weak value of $\overline{A_w} = \frac{1}{\epsilon}$, which has no improvement than using $N$ uncorrelated photons.

\subsection*{\uppercase\expandafter{\romannumeral3.C}. Fisher Information Analysis}

In order to determine how well an unknown parameter can be estimated, Fisher information is applied to describe the maximum available information one can achieve. The Fisher information is defined as\cite{Helstrom1976}:
\begin{equation}
I(g)=\int P(p|g)[\partial_g\log P(p|g)]^2 dx,
\end{equation}
where $g$ is the parameter of interest, and $P(p|g)$ is the conditional probability distribution of $p$ with a given $g$. In our case, for $\nu$ rounds of successful coincidence detections, and the total Fisher information is $\nu I(g)$. The lower bound of statistical error of $g$ is given by the Cram\'{e}r-Rao bound\cite{Helstrom1976}, which yields $\Delta g \geq \sqrt{\nu\cdot I^{-1}(g)}$.

When the $N$-order correlation function is used in estimating g by eqs.(\ref{eq:GNent}), the corresponding Fisher information can be computed by
\begin{equation}\label{eq:infor}
\begin{aligned}
I(g)= \int D_{\{p_n\}}G^{(N)}(\{p_n\})[\partial_g \log{G^{(N)}(\{p_n\})}]^2,
\end{aligned}
\end{equation}
where $D_{\{p_n\}}$ denotes $dp_1dp_2\cdots dp_N$. We then obtain the Fisher information for $\hat{M}=\hat{X}$ and $\hat{M}=\hat{P}$ with in a successful coincidence detection by:
\begin{subequations}\label{eq:main}
\begin{align}
&I_{\hat{X}}(g)=p_s(\mathrm{Re}A_w)^2/\sigma_0^2,\\
&I_{\hat{P}}(g)=4p_s(\mathrm{Im}\overline{A_w})^2\sigma_0^2.
\end{align}
\end{subequations}

One may argue that the amplification effect induced by $\mathrm{Re}A_w$ and $\mathrm{Im}\overline{A_w}$ in Eq.(\ref{eq:main}) may be eliminated by the low postselection probability\cite{Jordan2014}. Fortunately, even with low success probability, the Heisenberg scaling is still achievable by cleverly choosing initial/final-selection states and $\hat{M}$.
Adopting the initial and final selections in Eq.(\ref{eq:i-f}), we have $p_s \approx \epsilon^2$, $A_w \approx N/\epsilon$ and $\overline{A_w} = 1/\epsilon$. In this case,
for $\hat{M} = \hat{X}$ we obtain $I_{\hat{X}}(g)=N^2/\sigma_0^2$, and therefore $\Delta g_{min} \propto \frac{1}{\sqrt{\nu}N}$, which is the Heisenberg scaling precision one expects in quantum metrology\cite{Giovannetti2004}.


\section*{\uppercase\expandafter{\romannumeral4}. Discussions and Conclusion}
The real part of the weak value provides a measurable window to nonclassical features of quantum mechanics, as it can be interpreted as a conditioned average correlated to an observable\cite{Dressel2014}. However in previous experimental studies in WVA without entanglement\cite{Dixon2009,Xu2013}, imaginary weak value is preferred as it can achieve higher amplification. Interestingly in our scenario, Eqs.(\ref{eq:main}) show that it is the real part of the weak value, rather than its imaginary part, helps increasing the amplification factor and Fisher information.

Our proposal can be implemented by optic experiments in either spatial-momentum domain\cite{Ritchie1991} or time-frequency domain\cite{Brunner2010}. For the former case, one can use position imaging lens plus charge-coupled device(CCD) to measure the position correlation, or momentum imaging lens plus CCD to measure the momentum correlation\cite{Dressel2014,Pearce2015}. For the later case, time-correlation measurement\cite{Boitier2010} and spectral-correlation measurement\cite{Gerrits2015} techniques can be applied. Similar to Ref.\cite{Pang2014}, our proposal faces the same difficulty in selecting entangled postselection states. Some existing arrangements, such as Bell-state or GHZ-state analyzer\cite{Pan2012}, may be utilized in the experimental design.

In summary, we have revealed the relation between field correlation functions and weak measurement, and proposed a new way to detect the anomalous weak value amplification phenomenon. Different from the previous proposal\cite{Pang2014}, our scheme is implemented by hyperentangled photons along with coincidence detections, and optimally increase the amplification factor by a factor of $N$. For the purpose of parameter estimation, our scheme can achieve the Heisenberg limit, which is expected in quantum metrology.
Hyperentanglement has been shown useful in quantum information process\cite{Walborn2006} and quantum communication\cite{Sheng2010}, however its application in weak measurement and quantum metrology is not clear. By proposing a concrete hyperentanglement-enhanced WVA scheme, we fill in this blank and expect a further investigation.

\begin{acknowledgments}
This work was supported by the National Natural Science Foundation of China (Grants No. 61170228£¬No. 61332019£¬ 61471239), and the Hi-Tech Research and Development Program of China (Grant No: 2013AA122901).
\end{acknowledgments}

\bibliographystyle{unsrt}

\begin{thebibliography}{}
\bibitem{Note} In particular when optic configurations are in used, according to the relations $x = ct$ and $p = \hbar\omega/c$, the position mode and momentum mode correspond to the time domain and the optic frequency domain of the photons respectively.
\end{thebibliography}

\end{document}